\documentclass[aps,pre,showpacs,twocolumn,unsortedaddress]{revtex4-1}
\usepackage{graphicx}
\usepackage{dcolumn}
\usepackage{bm}
\usepackage{amsmath}
\usepackage{mathrsfs}
\usepackage{comment}
\usepackage{color}
\begin{document}

\title{A general formulation of long-range degree correlations in complex networks}

\author{Yuka Fujiki}
\email{y-fujiki@eng.hokudai.ac.jp}
\affiliation{Department of Applied Physics, Hokkaido University, Sapporo 060-8628, Japan}
\author{Taro Takaguchi}
\email{taro.takaguchi.cp@gmail.com}
\affiliation{National Institute of Information and Communications Technology, Tokyo 184-8795, Japan}
\author{Kousuke Yakubo}
\email{yakubo@eng.hokudai.ac.jp}
\affiliation{Department of Applied Physics, Hokkaido University, Sapporo 060-8628, Japan}

\date{\today}

\begin{abstract}
We provide a general framework for analyzing degree
correlations between nodes separated by more than one step
(i.e., beyond nearest neighbors) in complex networks. One
probability and four conditional probabilities are introduced
to fully describe long-range degree correlations with respect
to $k$ and $k'$ of two nodes and shortest path length $l$
between them. We present general relations among these
probabilities and clarify the relevance to nearest-neighbor
degree correlations. Unlike nearest-neighbor correlations, some
of these probabilities are meaningful only in finite-size
networks. Furthermore, as a baseline to determine the existence
or nonexistence of long-range degree correlations in a network,
the functional forms of these probabilities for networks
without any long-range degree correlations are analytically
evaluated within a mean-field approximation. The validity of
our argument is demonstrated by applying it to real-world
networks.
\end{abstract}
\pacs{89.75.Hc, 89.75.Fb, 02.70.Rr} \maketitle

\section{Introduction}
\label{sec:intro}
In many networks describing complex real systems, the number of
edges from a node, namely degree, widely fluctuates from node
to node, and degree distributions often exhibit power-law
behavior \cite{Barabasi99}. For such networks, significant
interest now concentrates on the issue of correlations between
degrees of two nodes. In particular, degree correlations
between adjacent nodes have been extensively studied so far
\cite{Pastor-Satorras01,
Maslov02,Newman02,Newman03,Park03,Catanzaro04,Serrano06,
Fotouhi13,Litvak13}. Nearest neighbor degree correlations
(NNDCs) in complex networks are related to their fundamental
structural properties, such as clustering
\cite{Soffer05,Serrano05, Miller09,Hofstad17}, community
structures \cite{Menche10}, the average path length
\cite{Xulvi-Brunet04}, and fractality
\cite{Yook05,Song06,Wei16}. In addition, NNDCs influence
various dynamics on networks, such as epidemic spreading
\cite{Eguiluz02,Boguna02,Boguna03,Gross06, Hindes17},
synchronization phenomena \cite{Sorrentino06,Bernardo07,
LaMar10,LaRocca11,Avalos-Gaytan12,Jalan16}, strategic games
\cite{Rong07,Rong09,Devlin09,La17}, and resilience to failures
\cite{Noh07,Goltsev08,Schneider11,Tanizawa12,Watanabe16}.

It has, however, been pointed out recently that NNDCs are not
enough to characterize structural properties of complex
networks. For example, scale-free fractal networks are known to
exhibit negative NNDCs (namely, disassortative mixing)
\cite{Yook05}. Thus, hub nodes in such a network are almost
never connected directly by an edge. In actual fractal
networks, like the World Wide Web or synthetic graphs
\cite{Song06,Rozenfeld07}, however, hub nodes are not only
nonadjacent to, but also repulsive over a long-range distance
to each other \cite{Fujiki17}. As another example,
Orsini \textit{et al.}~\cite{Orsini15} found that many local
and even global structural features of real-world complex networks
are closely reproduced by random graphs with the same degree
sequences, clustering, and NNDCs as those for the real networks.
However, some sort of global properties, such as the shortest path
length distributions, betweenness distributions, and community
structures, cannot be explained by these local characteristics.
This implies that intrinsic non-local degree correlations in
these networks cannot be described by NNDCs as a local
characteristic. Furthermore, it has been demonstrated that the
shortest path length between hub nodes influences functions or
dynamical properties of networks
\cite{Tadic04,Boguna13,Boulos13,Swanson16}. For understanding
non-local structural properties, it is important and useful to
provide a framework to describe degree correlations between
nodes beyond nearest neighbors, namely, long-range degree
correlations (LRDCs).

There have been several proposals for formulating LRDCs in
complex networks. Rybski \textit{et al.}~\cite{Rybski10}
describe LRDCs by fluctuations of the degree along shortest
paths between two nodes. This is an analogy to fluctuation
analysis used in correlated time series.
Mayo \textit{et al.}~\cite{Mayo15} defined the long-range
assortativity and the average $l$th neighbor degree to quantify
LRDCs (the same definition of the long-range assortativity was
independently employed in \cite{Arcagni17}). The long-range
assortativity $r_{l}$ is the Pearson correlation coefficient
between degrees of pairs of nodes separated by the shortest
path length $l$ from each other. The average $l$th neighbor
degree $k_{l}(k)$ is the average degree of nodes separated by
$l$ from a node of degree $k$. They found that social networks
exhibit disassortative degree correlations on long-range
scales, while nonsocial networks do not indicate such a
tendency. The two-walks degree assortativity proposed by
Allen-Perkins \textit{et al.}~\cite{Allen-Perkins17} is another
type of assortativity measure beyond nearest neighbors. This
quantity is defined as the Pearson correlation coefficient of
the sum of the nearest-neighbor degrees of adjacent nodes,
which reflects second neighbor degree correlations. These
quantities enable us to pick up some specific aspects of LRDCs.
However, if we perform a global and multilateral analysis of
LRDCs, a more general framework is required to obtain the
entire information of LRDCs.

In this work, we provide a general framework for analyzing
LRDCs in complex networks of either finite or infinite size. In
order to fully describe correlations between degrees $k$ and
$k'$ of two nodes separated by a shortest path length $l$, one
joint probability and four conditional probabilities are
introduced as functions of $k$, $k'$, and $l$. NNDCs can be
described by these probability functions as a special case of
$l=1$. These five probabilities are not independent of each
other, and we present general relations among them. In
addition, the functional forms of these probabilities for a
network without any LRDCs (referred as a long-range uncorrelated
network hereafter) are analytically evaluated within a
mean-field approximation. By comparing the probabilities for a
given network with those for the corresponding long-range
uncorrelated network, one can judge whether the network
possesses LRDCs or not, and obtain detailed information about
degree correlations. Finally, we demonstrate the validity of
our argument by applying it to real-world networks.

The rest of this paper is organized as follows. In
Sec.~\ref{sec:probabilities}, we introduce the probability functions
characterizing LRDCs and present general relations between them.
In Sec.~\ref{sec:uncorrelated}, the functional forms
of the probabilities for long-range uncorrelated networks are
analytically evaluated. In Sec.~\ref{sec:real}, the validity
of our argument is tested by calculating the probabilities for
real-world networks. Section \ref{sec:conclusion} is devoted to
the summary and remarks.

\section{Joint and conditional probabilities}
\label{sec:probabilities}
Degree correlations between nearest-neighbor nodes (namely,
NNDCs) are completely described by the joint probability
$P_{\text{nn}}(k,k')$ that two end nodes of a randomly chosen
edge have the degrees $k$ and $k'$. We can define the
conditional probability from $P_{\text{nn}}(k,k')$ by
$P_{\text{nn}}(k'|k)=P_{\text{nn}}(k,k')/\sum_{k'}P_{\text{nn}}(k,k')$,
which is the probability that a node adjacent to a randomly
chosen node of degree $k$ has the degree $k'$. If the degree
distribution function $P(k)$ is given, the probability
$P_{\text{nn}}(k'|k)$ also identifies NNDCs. We extend this
idea to LRDCs. All information pertaining to correlations
between degrees $k$ and $k'$ of two nodes separated by a
shortest path length $l$ (namely, LRDCs) is included in the
joint probability $P(k,k',l)$ that randomly chosen two nodes
have the degrees $k$ and $k'$ and the shortest path length
between them is $l$. From this joint probability, four
conditional probabilities can be constructed as follows,
\begin{subequations}
\begin{eqnarray}
P(l|k,k') &=& \frac{P(k,k',l)}{\sum_{l}P(k,k',l)}, \label{def_pl_kk} \\
P(k'|k,l) &=& \frac{P(k,k',l)}{\sum_{k'}P(k,k',l)}, \label{def_pk_kl} \\
P(k,k'|l) &=& \frac{P(k,k',l)}{\sum_{k,k'}P(k,k',l)}, \label{def_pkk_l} \\
P(k',l|k) &=& \frac{P(k,k',l)}{\sum_{k',l}P(k,k',l)}. \label{def_pkl_k}
\end{eqnarray}
\label{def_cond_p}
\end{subequations}
The meanings of these probabilities, as well as the joint
probability, are listed in Table \ref{table:1}. These
conditional probabilities also describe LRDCs. The
probabilities in Table \ref{table:1} are normalized as
$\sum_{k,k',l}P(k,k',l)=\sum_{l}P(l|k,k')=
\sum_{k'}P(k'|k,l)=\sum_{k,k'}P(k,k'|l)=\sum_{k',l}P(k',l|k)=1$.
Here, we note that the sum over $l$ includes the distance
($l_{\infty}$) between disconnected node pair. It should be
also emphasized that $P(k,k',l)$, $P(l|k,k')$, and $P(k',l|k)$
are meaningless for networks with infinitely large components
because values of these probabilities become always zero for
finite $l$. In contrast, $P(k'|k,l)$ and $P(k,k'|l)$ can be
properly defined even for infinite networks.
\begin{table}[ttt]
  \caption{\label{table:1}
Meanings of one joint probability and four conditional
probabilities characterizing LRDCs in networks.}
 \begin{ruledtabular}
 {\renewcommand\arraystretch{1.7}
  \begin{tabular}{lp{6.5cm}}
    Probability     &  Meaning \\ \hline
    $P(k,k',l)$     &  Probability that randomly chosen two nodes have the degrees $k$ and $k'$ and the distance between them is $l$ \\
    $P(l|k,k')$     &  Probability that randomly chosen two nodes of degrees $k$ and $k'$ are separated by $l$, namely,
the shortest path length distribution between nodes of degrees $k$ and $k'$ \\
    $P(k'|k,l)$     &  Probability that a node separated by $l$ from a randomly chosen node of degree $k$ has the degree $k'$, namely,
the degree distribution of a node separated by $l$ from a node of degree $k$ \\
    $P(k,k'|l)$     &  Probability that randomly chosen two nodes separated by $l$ from each other have the degrees $k$ and $k'$ \\
    $P(k',l|k)$     &  Probability that a randomly chosen node has the degree $k'$ and is separated by $l$ from a node of degree $k$
  \end{tabular}}
 \end{ruledtabular}
\end{table}

Using the joint probability $P(k,k',l)$, the degree distribution
$P(k)$ and the shortest path length distribution $R(l)$ are
presented by
\begin{equation}
P(k)=\sum_{k',l}P(k,k',l) ,
\label{pk}
\end{equation}
and
\begin{equation}
R(l)=\sum_{k,k'}P(k,k',l) ,
\label{rl}
\end{equation}
respectively. It is convenient to introduce the probability $Q(k|l)$
defined by
\begin{equation}
Q(k|l)=\sum_{k'}P(k,k'|l) ,
\label{qkl}
\end{equation}
which is the probability that one of two nodes separated by $l$
has the degree $k$. This is an extension of the probability
$Q_{\text{nn}}(k)$ that one end node of an \textit{edge} has
the degree $k$ to a long-range node pair in the sense of
$Q_{\text{nn}}(k)=Q(k|l=1)$. With the aid of $Q(k|l)$, we have
\begin{equation}
\sum_{k'}P(k,k',l)=Q(k|l)R(l) .
\label{sum_k_pkkl}
\end{equation}
Equations (\ref{pk}), (\ref{rl}), and (\ref{sum_k_pkkl}), as well
as the obvious relation
\begin{equation}
\sum_{l}P(k,k',l)=P(k)P(k') ,
\label{sum_l_pkkl}
\end{equation}
form sum rules of the joint probability $P(k,k',l)$. Considering
these sum rules, Eq.~(\ref{def_cond_p}) leads several general
relations between the conditional probabilities, $P(k)$, and
$R(l)$, such as,
\begin{gather}
P(k',l|k)=P(k')P(l|k,k') ,
\label{rel_1} \\
P(k,k'|l)=Q(k|l)P(k'|k,l) ,
\label{rel_2} \\
P(k)P(k',l|k)=Q(k'|l)R(l)P(k|k',l) ,
\label{rel_3}
\end{gather}
and
\begin{equation}
P(k)P(k')P(l|k,k')=R(l)P(k,k'|l) .
\label{rel_4}
\end{equation}
Equations (\ref{rel_3}) and (\ref{rel_4}) can be considered as
direct consequences of the Bayes' theorem that relates $P(A|B)$
and $P(B|A)$ for events $A$ and $B$.

The joint probability $P_{\text{nn}}(k,k')$ and the conditional
probability $P_{\text{nn}}(k'|k)$ describing NNDCs are included
in the above long-range probabilities as a special case of $l=1$.
In fact, we have
\begin{equation}
P_{\text{nn}}(k,k')=P(k,k'|l=1) ,
\label{rel_nnp_1}
\end{equation}
and
\begin{equation}
P_{\text{nn}}(k'|k)=P(k'|k,l=1) .
\label{rel_nnp_2}
\end{equation}
Similarly, the degree distribution $Q_{\text{nn}}(k)$ of an end node
of a randomly chosen edge is given by
\begin{equation}
Q_{\text{nn}}(k)=Q(k|l=1)=\frac{kP(k)}{\langle k\rangle} ,
\label{rel_nnp_3}
\end{equation}
where $\langle k\rangle=\sum_{k}kP(k)$ is the average degree.
Then, Eq.~(\ref{rel_2}) with $l=1$ is reduced to the well-known
relation $P_{\text{nn}}(k,k')=kP(k)P_{\text{nn}}(k'|k)/\langle k\rangle$
\cite{Boguna02}.

Considering the above correspondence, we can easily extend indices
characterizing NNDCs to those for LRDCs. For example, the long-range
assortativity $r_{l}$ can be defined as
\begin{equation}
r_{l}=\frac{4\langle kk'\rangle_{l}-\langle k+k' \rangle_{l}^{2}}
{2\langle k^{2}+k'^{2}\rangle_{l}-\langle k+k' \rangle_{l}^{2}} ,
\label{l_assort}
\end{equation}
where $\langle f(k,k')\rangle_{l}=\sum_{k,k'}f(k,k')P(k,k'|l)$.
This quantity is the Pearson correlation coefficient between
degrees of nodes separated by $l$ from each other. For $l=1$,
$r_{l}$ is reduced to the conventional nearest-neighbor
assortativity\cite{Newman02}. Another example is the average
degree of $l$th neighbor nodes, which is given by \cite{comment1}
\begin{equation}
k_{l}(k)=\sum_{k'}k'P(k'|k,l) .
\label{klk}
\end{equation}
This is an extension of the average degree, $k_{\text{nn}}(k)=
\sum_{k'}k'P_{\text{nn}}(k'|k)$, of nearest neighbors of a node
of degree $k$  to that for $l$th neighbors. The quantities $r_{l}$
and $k_{l}(k)$ are equivalent to those proposed by Ref.~\cite{Mayo15}.
Besides extensions of existing indices for NNDCs, it is also
possible to introduce completely new measures characterizing
LRDCs, such as the strength of long-range repulsive correlations
between hubs, by using the probabilities listed in Table \ref{table:1}.

\section{Long-range uncorrelated networks}
\label{sec:uncorrelated}
In the previous section, we introduced five fundamental
probabilities describing LRDCs in complex networks. However,
even if we know these probabilities for a given network, we
cannot judge whether the network possesses LRDCs or not. This
is due to the lack of a baseline for comparison, i.e., it has
not yet been clarified how these probabilities behave for a
network in which the degrees of two nodes separated by an
arbitrary distance are not correlated. In this section, we
evaluate functional forms of the probabilities for long-range
uncorrelated networks (LRUNs).

\subsection{General remarks}
\label{subsec:remarks}
A nearest-neighbor uncorrelated network (NNUN) is a network
in which the degree of one end node of an edge is independent
of the degree of another end node. Thus, the joint probability
$P_{\text{nn}}(k,k')$ in an NNUN is given by the product
$Q_{\text{nn}}(k)Q_{\text{nn}}(k')$. Extending this idea, an LRUN
is considered to be a network satisfying the relation
\begin{equation}
P_{0}(k,k'|l)=Q_{0}(k|l)Q_{0}(k'|l) ,
\label{def_uncorrelated}
\end{equation}
for any $l$, where $P_{0}(k,k'|l)$ and $Q_{0}(k|l)$ represent
$P(k,k'|l)$ and $Q(k|l)$ for LRUNs, respectively. Hereafter, we
denote the probabilities for LRUNs by adding the subscript
``$0$". Equation (\ref{def_uncorrelated}) implies that the
degrees $k$ and $k'$ of two nodes separated by $l$ are
independent of each other.

While $P_{\text{nn}}(k,k')$ for an NNUN has the simple functional
form as $P_{\text{nn}}(k,k')=kk'P(k)P(k')/\langle k \rangle^{2}$,
it is difficult to obtain an exact expression of $P_{0}(k,k'|l)$
because $Q(k|l)$ itself depends on LRDCs. Nevertheless, we can
generally conclude that $P_{0}(k'|k,l)$ for an LRUN does not depend
on $k$. This comes immediately from the relation
\begin{equation}
P_{0}(k'|k,l)=Q_{0}(k'|l) ,
\label{uc_rel_1}
\end{equation}
which is obtained by substituting Eq.~(\ref{def_uncorrelated})
into Eq.~(\ref{rel_2}). Equation (\ref{uc_rel_1}) with $l=1$
leads the well-known relation $P_{\text{nn}}(k'|k)=k'P(k')/
\langle k\rangle$ for NNUNs \cite{Boguna02}.

\subsection{Mean-field approximation}
\label{subsec:mean-field}
We have mentioned that the probability functions for LRUNs are
not easy to calculate rigorously. A major reason for this
difficulty is that finite sizes of networks are essential for
these probabilities, as pointed out in
Sec.~\ref{sec:probabilities}. Thus, we need to approximate
these probabilities for LRUNs. Once one of these probabilities
is obtained, other probabilities can be calculated by using
Eq.~(\ref{def_cond_p}). We then focus on $P_{0}(l|k,k')$ at
first, which is the length distribution between nodes of
degrees $k$ and $k'$ and has been argued recently by Melnik and
Gleeson \cite{Melnik16}. They calculated $P(l|k,k')$ for finite
random networks such as Erd\H{o}s-R\'{e}nyi random graphs or
networks generated by the configuration model \cite{Newman01}. We
can reasonably assume that finite random networks belong to the
class of LRUNs when their sizes are finite but sufficiently large,
from the fact that infinite random networks satisfy
Eq.~(\ref{def_uncorrelated}) to be LRUNs. Therefore, we take
$P(l|k,k')$ in their calculation \cite{Melnik16} as
$P_{0}(l|k,k')$ for LRUNs, after some necessary modifications
\cite{comment2}.

Let us introduce the probability $\rho(l|k,k')$ that the
distance between randomly chosen two nodes of degrees $k$ and
$k'$ is equal to or less than $l$. The probability
$P_{0}(l|k,k')$ is then presented by
\begin{equation}
P_{0}(l|k,k')=
\begin{cases}
\rho(0|k,k')                                    & (l=0) ,    \\[3pt]
\rho(l|k,k')-\rho(l-1|k,k')                     & (l\ge 1) , \\[3pt]
1-\displaystyle \lim_{l\to \infty}\rho(l|k,k')  & (l=l_{\infty}) .
\end{cases}
\label{p_lkk_2}
\end{equation}
The last expression for $l=l_{\infty}$ is for disconnected node
pairs in a network composed of multiple components. The
normalization condition of $P(l|k,k')$ is thus written as
$\sum_{l=0}^{\infty}P(l|k,k')+P(l_{\infty}|k,k')=1$.

Under the local tree assumption and the mean-field approximation,
$\rho(l|k,k')$ for a random network is given by \cite{Melnik16}
\begin{equation}
\rho(l|k,k')=1-\left[1-\rho(0|k,k')\right]\left[1-\bar{q}(l-1|k,k')\right]^{k} ,
\label{cal_rho}
\end{equation}
where $\bar{q}(l|k,k')$ is the probability that an adjacent
node of a randomly chosen node $i_{k}$ of degree $k$ lies
within the distance $l$ from a node $j_{k'}$ of degree $k'$
under the condition that $i_{k}$ is separated by more than
$l$ from $j_{k'}$. The first factor $1-\rho(0|k,k')$ of the
second term in the right-hand side represents the probability
that the node $i_{k}$ is not the node $j_{k'}$ itself. The
second factor
$\left[1-\bar{q}(l-1|k,k')\right]^{k}$ means the probability
that all adjacent nodes of $i_{k}$ are separated by more than
$l-1$ from $j_{k'}$ under the condition that $i_{k}$ is
separated by more than $l-1$ from $j_{k'}$. Thus, the rough
meaning of Eq.~(\ref{cal_rho}) is that the probability that
the node $i_{k}$ lies within the distance $l$ from the node
$j_{k'}$ is equal to the probability that at least one of
$k$ adjacent nodes of $i_{k}$ lies within the distance $l-1$
from $j_{k'}$. Furthermore, let us introduce the probability
$q(l|k,k')$ that a randomly chosen node $i_{k}$ of degree
$k$ with at least one neighboring node, say $h$, separated
by more than $l$ from a node $j_{k'}$ of degree $k'$ lies
within the distance $l$. Then, we have the following relation
between $q(l|k,k')$ and $\bar{q}(l-1|k,k')$ similar to
Eq.~(\ref{cal_rho}),
\begin{equation}
q(l|k,k')=1-\left[1-\rho(0|k,k')\right]\left[1-\bar{q}(l-1|k,k')\right]^{k-1} .
\label{rel_qbar_q_1}
\end{equation}
The right-hand side of this equation implies the probability
that at least one node of $k-1$ adjacent nodes of $i_{k}$
other than $h$ lies within the distance $l-1$ from $j_{k'}$.
Using $q(l|k,k')$, the probability $\bar{q}(l|k,k')$ is
expressed by
\begin{eqnarray}
\bar{q}(l|k,k')&=&\sum_{k''}P_{\text{nn}}(k''|k)q(l|k'',k') , \nonumber \\
&=&\frac{1}{\langle k\rangle}\sum_{k''}k''P(k'')q(l|k'',k') .
\label{rel_qbar_q_2}
\end{eqnarray}
Since $\bar{q}(l|k,k')$ is actually independent of $k$,
we denote it simply by $\bar{q}(l|k')$. Multiplying
$kP(k)/\langle k\rangle$ on both sides of Eq.~(\ref{rel_qbar_q_1}),
summing over $k$, and using Eq.~(\ref{rel_qbar_q_2}), we have
the recursion equation for $\bar{q}(l|k')$,
\begin{eqnarray}
\bar{q}(l|k')=1&-&G_{1}\left[1-\bar{q}(l-1|k')\right] \nonumber \\
&+&\frac{k'}{N\langle k\rangle}\left[1-\bar{q}(l-1|k')\right]^{k'-1} ,
\label{recursion_eq}
\end{eqnarray}
where $N$ is the number of nodes in the network and $G_{1}(x)$ is
the generating function defined by
$G_{1}(x)=\sum_{k}x^{k-1}kP(k)/\langle k\rangle$. Here, we used
the obvious relation,
\begin{equation}
\rho(0|k,k')=\frac{\delta_{kk'}}{NP(k)} .
\label{ini_cond1}
\end{equation}
Equation (\ref{recursion_eq}) can be solved iteratively with the
initial condition \cite{Melnik16},
\begin{equation}
\bar{q}(0|k')=\frac{k'}{N\langle k\rangle} .
\label{ini_cond2}
\end{equation}
Using the solution of $\bar{q}(l|k')$ and Eqs.~(\ref{p_lkk_2})
and (\ref{cal_rho}), we can calculate $P_{0}(l|k,k')$. The
joint probability $P_{0}(k,k',l)$ is computed by
$P(k)P(k')P_{0}(l|k,k')$ from Eqs.~(\ref{def_pl_kk}) and
(\ref{sum_l_pkkl}), and other conditional probabilities listed
in Table \ref{table:1} are determined from $P_{0}(k,k',l)$ by
using Eq.~(\ref{def_cond_p}).

We should remark the accuracy of the mean-field approximation
in the above calculation. The probability $\rho(l|k,k')$ must
be equal to $\rho(l|k',k)$ from the definition. However,
$\rho(l|k,k')$ calculated from Eq.~(\ref{cal_rho}) is actually
not symmetric with respect to $k$ and $k'$. In fact, $\rho(l|k,k')$
for $l=1$ and $k\ne k'$, calculated as
\begin{eqnarray}
\rho(1|k,k') &=& 1-\left(1-\frac{k'}{N\langle k\rangle}\right)^{k} \nonumber \\
&=& \frac{kk'}{N\langle k\rangle} -\frac{1}{2}\frac{k(k-1)k'^{2}}{\left(N\langle k\rangle\right)^{2}}+\cdots ,
\end{eqnarray}
is asymmetric in the order of $N^{-2}$. This is due to
the difference in accuracy of the mean-field treatment for
nearest neighbors of the nodes of degrees $k$ and $k'$.
The mean-field approximation for neighboring nodes of a large
degree node is more accurate than that of a small degree
node. Since $\bar{q}(l|k')$ is iteratively calculated for the
distance $l$ from the source node of degree $k'$ according to
Eq.~(\ref{recursion_eq}), $\rho(l|k,k')$ with $k<k'$ is more
accurate than $\rho(l|k',k)$. Therefore, we first calculate
$\rho(l|k,k')$ for $k<k'$ by Eq.~(\ref{cal_rho}), then transfer
it to $\rho(l|k',k)$ in actual computations.
Another remark on the mean-field approximation is related to
the component-size distribution. We assume that $\rho(l|k,k')$
does not depend on the size of the component that
the source node of degree $k'$ belongs to. This implies that
the distribution function of the component size is assumed to
be relatively narrow. If a random network with a given degree
distribution $P(k)$ is very close to its percolation transition
point, however, the component-size distribution becomes wide,
and then the mean-field calculations have poor accuracy.

\subsection{Infinite tree-like networks}
\label{subsec:infinite}
For infinitely large networks, only $P(k'|k,l)$ and $P(k,k'|l)$
are meaningful among five probabilities, as mentioned in
Sec.~\ref{sec:probabilities}. It is easy to calculate these
conditional probabilities for infinite random networks with
tree-like structures. Let us consider $P_{0}(k'|k,l)$ at first.
Since this is the probability that a node separated by $l$
from a node of degree $k$ has the degree $k'$, $P_{0}(k'|k,l)$
must satisfy the relation,
\begin{equation}
P_{0}(k'|k,l)=\sum_{k''}P_{\text{nn}}(k'|k'')P_{0}(k''|k,l-1) ,
\label{inf_pk_kl_1}
\end{equation}
where the nearest-neighbor degree distribution function
$P_{\text{nn}}(k'|k'')$ is given by $k'P(k')/\langle k\rangle$
for random networks. Using the obvious relation
$P_{0}(k'|k,0)=\delta_{kk'}$, we can solve the above equation
as,
\begin{equation}
P_{0}(k'|k,l)=
\begin{cases}
\delta_{kk'}                                   &  (l=0) ,    \\[5pt]
\displaystyle \frac{k'P(k')}{\langle k\rangle} & (l\ge 1) .
\end{cases}
\label{inf_pk_kl_2}
\end{equation}
Thus, we have immediately, from Eq.~(\ref{uc_rel_1}),
\begin{equation}
Q_{0}(k|l)=\frac{kP(k)}{\langle k\rangle} \quad (l\ge 1) .
\label{q0_k_l}
\end{equation}
The probability $P_{0}(k,k'|l)$ for $l\ge 1$ is then calculated
from Eq.~(\ref{rel_2}) as
\begin{equation}
P_{0}(k,k'|l)=\frac{kk'P(k)P(k')}{\langle k\rangle^{2}} .
\label{p0_kk_l}
\end{equation}
We should note that $P_{0}(k,k'|l)$ and $P_{0}(k'|k,l)$ for infinite
tree-like random networks are equivalent to $P_{\text{nn}}(k,k')$
and $P_{\text{nn}}(k'|k)$, respectively, independently of $l$.

It is reasonable to consider that the above expressions of
$P_{0}(k'|k,l)$, $Q_{0}(k|l)$, and $P_{0}(k,k'|l)$ for infinitely
large networks hold approximately for $l\ll \langle l\rangle$ even
in finite random networks, where $\langle l\rangle$ is the average
shortest path length. While we have shown in Sec.~\ref{subsec:remarks}
that $P_{0}(k'|k,l)$, in general, does not depend on $k$, our
result here indicates that this probability is independent of
$l$ too if $l\ll \langle l\rangle$.

\begin{figure*}[ttt!]
\begin{center}
\includegraphics[width=0.95\textwidth]{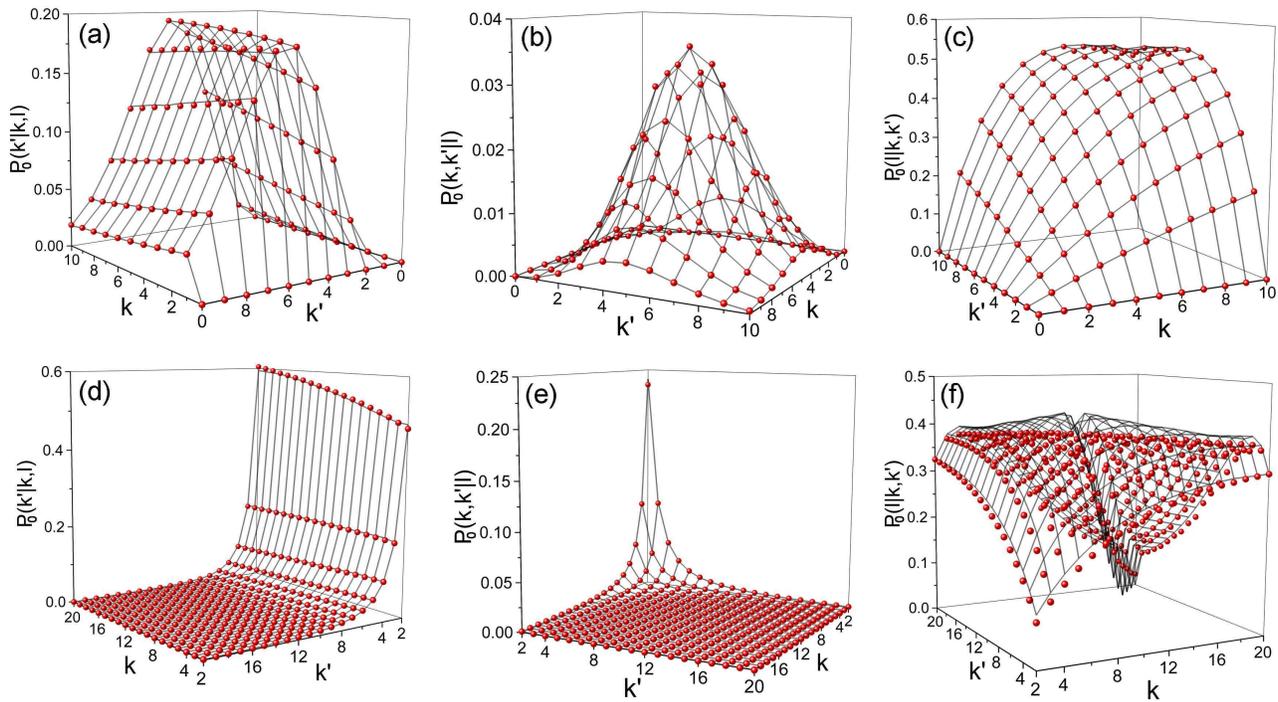}
\caption{(Color online) Probabilities $P_{0}(k'|k,l)$, $P_{0}(k,k'|l)$,
and $P_{0}(l|k,k')$ as functions of $k$ and $k'$ for $l=4$ and $N=1,000$.
Wireframes indicate the probabilities calculated analytically by the
method explained in Sec.~\ref{subsec:mean-field}, while dots represent
those measured for numerically realized random networks. Upper three
panels [(a)-(c)] are the results for Erd\H{o}s-R\'{e}nyi random graphs
with $\langle k\rangle=5.0$, and lower panels [(d)-(f)] for
scale-free random networks with the degree distribution
$P(k)\propto k^{-3}$.}
\label{fig:1}
\end{center}
\end{figure*}
\subsection{Numerical confirmation}
\label{subsec:numerical}
In order to confirm the validity of our analytical evaluation of
the probability functions for LRUNs, we compare the probabilities
$P_{0}(l|k,k')$, $P_{0}(k'|k,l)$, and $P_{0}(k,k'|l)$ obtained by
the method explained in Sec.~\ref{subsec:mean-field} with those
measured for synthetic random networks. Figure 1 shows
the dependence of these probabilities on $k$ and $k'$ for $l=4$. The
wireframe in each panel indicates the analytically calculated
probabilities, while dots represent numerical results. The upper
three panels give the results for Erd\H{o}s-R\'{e}nyi random graphs
with $\langle k\rangle=5.0$ and $N=1,000$. We have dared to employ
relatively small networks to check the validity of the method for
finite sizes. Numerical results are obtained by averaging over $100$
realizations of Erd\H{o}s-R\'{e}nyi random graphs. The average path
length of these networks is $\langle l\rangle=4.5$. The lower three
panels present the results for scale-free random networks with
$N=1,000$ and the degree distribution function of $P(k)\propto k^{-3}$
for $2\le k\le 50$ and $P(k)=0$ otherwise. Numerical results show
the averages over $10,000$ realizations generated by the configuration
model. The average degree and the average path length are
$\langle k\rangle=3.1$ and $\langle l\rangle=5.4$, respectively.
These plots demonstrate that the analytical treatment based on
the mean-field approximation well reproduces numerical results
even for finite networks.

\begin{figure}[bbb!]
\begin{center}
\includegraphics[width=0.35\textwidth]{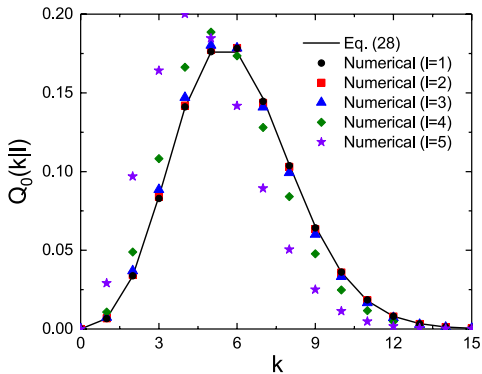}
\caption{(Color online) Probability $Q_{0}(k|l)$ as a function of
$k$ for Erd\H{o}s-R\'{e}nyi random graphs with $N=1,000$ and
$\langle k\rangle=5.0$. Solid line indicates the probability given
by Eq.~(\ref{q0_k_l}), while symbols represent the numerical results
for $l=1$ (circles), $2$ (boxes),$3$ (triangles), $4$ (diamonds),
and $5$ (stars) averaged over $100$ network realizations.}
\label{fig:2}
\end{center}
\end{figure}
We also verified the argument in Sec.~\ref{subsec:infinite} by
calculating $Q_{0}(k|l)$ for Erd\H{o}s-R\'{e}nyi random graphs.
Figure \ref{fig:2} compares this probability calculated by
Eq.~(\ref{q0_k_l}) with $Q_{0}(k|l)$ measured numerically.
The Erd\H{o}s-R\'{e}nyi random graphs are the same as in
Fig.~\ref{fig:1}. Thus, the average path length is
$\langle l\rangle=4.5$ for these random graphs. As shown by
the numerical results for $l=1$, $2$, and $3$, $Q_{0}(k|l)$
measured numerically is almost independent of $l$ and is well
described by Eq.~(\ref{q0_k_l}), if $l$ is sufficiently smaller
than $\langle l\rangle$. On the contrary, if $l$ becomes close
to or larger than $\langle l\rangle$, numerically computed
$Q_{0}(k|l)$ deviates from Eq.~(\ref{q0_k_l}), as shown by
the results for $l=4$ and $5$. These results prove
that Eq.~(\ref{q0_k_l}) holds for $l\ll \langle l\rangle$ even
in finite networks.

\section{Real-world networks}
\label{sec:real}
Finally, we investigate LRDCs in two real-world complex
networks by using the probabilities listed in Table
\ref{table:1}. One is the Gnutella peer-to-peer network
\cite{gnutella} and the other is the coauthorship network
\cite{coauthorship}. The Gnutella network has
$N=10,876$ nodes and $E=39,994$ edges. Thus, the average degree
is $\langle{k}\rangle=7.4$. This network consists of a single
connected component. The average path length $\langle l\rangle$
and the maximum shortest path length (diameter)
$l_{\text{max}}$ are $4.6$ and $10$, respectively. The
Spearman's degree-rank correlation coefficient $\rho$
\cite{Litvak13,Zhang16} characterizing the NNDC is measured as
$\rho=0.0$, which implies no NNDC in the Gnutella network.
The coauthorship network possesses $N=23,133$ nodes and
$E=93,439$ edges, which give $\langle k\rangle=8.1$. This
network is composed of the largest connected component with
$21,363$ nodes and $566$ small components with $3.1$ nodes on
average. The average path length and the network diameter are
$\langle l\rangle=5.4$ and $l_{\text{max}}=15$, respectively.
The Spearman's correlation coefficient of the coauthorship
network is $\rho=0.26$, which means a positive NNDC.

\begin{figure}[ttt!]
\begin{center}
\includegraphics[width=0.48\textwidth]{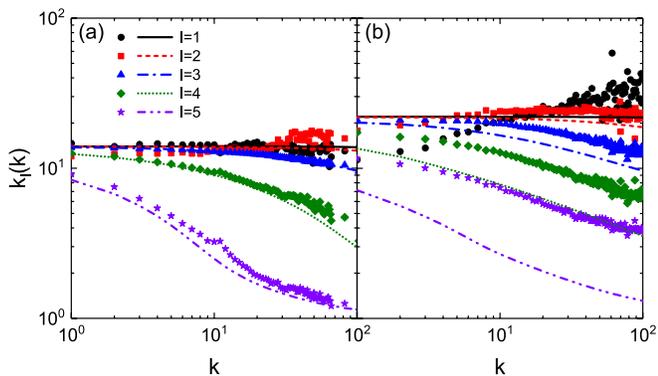}
\caption{(Color online) Average $l$th neighbor degree $k_{l}(k)$ of
a typical node of degree $k$ for (a) the Gnutella network \cite{gnutella}
and (b) the coauthorship network \cite{coauthorship}. Symbols represent
$k_{l}(k)$ of the real-world networks as a function of $k$ at fixed
values of $l$, and curves indicate those of corresponding LRUNs
with the same degree sequences as the real networks.}
\label{fig:3}
\end{center}
\end{figure}
For these two real-world networks, we first calculate the
average $l$th neighbor degree $k_{l}(k)$ given by
Eq.~(\ref{klk}). The results are presented by symbols in
Fig.~\ref{fig:3}. The continuous curves in this figure indicate
$k_{l}(k)$ for LRUNs with the same degree sequences as the
real networks, which is calculated from Eq.~(\ref{klk}) by
replacing $P(k'|k,l)$ with $P_{0}(k'|k,l)$. The symbols for
various $l$ in Fig.~\ref{fig:3}(a) are approximately fitted by
the corresponding curves. This implies that the Gnutella
network has almost no LRDCs. On the contrary, $k_{l}(k)$ for
the coauthorship network [Fig.~\ref{fig:3}(b)] considerably
deviates from the curves, and the discrepancy becomes more
pronounced at the higher degrees. This result clearly
demonstrates the LRDC in the coauthorship network in which the
average $l$th neighbor degree is always larger than that
expected for the LRUNs.

\begin{figure}[ttt!]
\begin{center}
\includegraphics[width=0.35\textwidth]{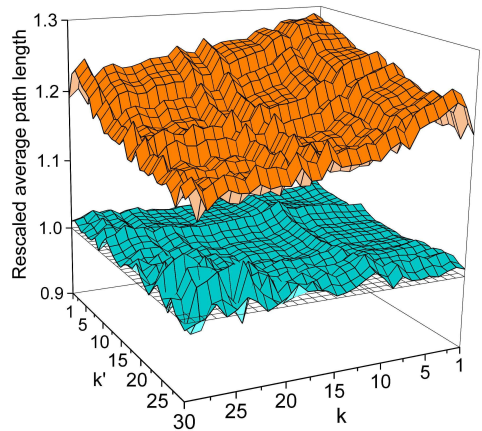}
\caption{(Color online) Rescaled average shortest path length
$\langle l(k,k')\rangle_{\text{res}}$ between nodes of degree
$k$ and $k'$. The lower and upper surfaces represent the results
for the Gnutella network and the coauthorship network,
respectively. The bottom flat mesh indicates
$\langle l(k,k')\rangle_{\text{res}}=1$.}
\label{fig:4}
\end{center}
\end{figure}

We also evaluate, for these two networks, the average shortest
path length $\langle l(k,k')\rangle$ between nodes of degrees
$k$ and $k'$, which is defined by
\begin{equation}
\langle l(k,k')\rangle = \sum_{l}lP(l|k,k') .
\label{lavkk}
\end{equation}
Figure \ref{fig:4} represents the results for the Gnutella and
the coauthorship networks. The vertical axis indicates the
average shortest path length rescaled by that for LRUNs with
the same degree sequence, namely,
$\langle l(k,k')\rangle_{\text{res}}=\sum_{l}lP(l|k,k')/\sum_{l}lP_{0}(l|k,k')$.
Although the maximum degrees $k_{\text{max}}$ of these networks
are larger than the range of $k$ in Fig.~\ref{fig:4}
($k_{\text{max}}=103$ for the Gnutella network and $279$ for
the coauthorship network), we depict the results only for
$k,k'\le 30$, in which $99.1$\% and $96.6$\% of nodes in the
Gnutella network and the coauthorship network are included,
respectively. This is because $\langle
l(k,k')\rangle_{\text{res}}$ for large degrees becomes quite
bumpy due to poor statistics by the less number of high degree
nodes. We see from Fig.~\ref{fig:4} that $\langle
l(k,k')\rangle_{\text{res}}$ for the Gnutella network is close
to $1$ independently of $k$ and $k'$. This means that the
network has almost no LRDCs, which is consistent with the
result shown in Fig.~\ref{fig:3}(a). In contrast, $\langle
l(k,k')\rangle_{\text{res}}$ for the coauthorship network is
larger than unity. This clearly indicates repulsive
correlations among nodes. In fact, the average path length
$\langle l\rangle=5.4$ for the coauthorship network is greater
than $\langle l\rangle=4.3$ for LRUNs with the same degree
sequence, whereas $\langle l\rangle=4.6$ for the Gnutella
network does not change so much from $\langle l\rangle=4.5$ for
the corresponding LRUNs. The fact that, for the coauthorship
network, $\langle l(k,k')\rangle_{\text{res}}$ for small
degrees is larger than that for large degrees demonstrates the
LRDC in which small degree nodes strongly repel each other in
this network.

\section{Conclusions}
\label{sec:conclusion}
In this paper, we have provided a general framework to analyze
pairwise correlations between degrees of nodes at an arbitrary
distance from each other in a complex network. In order to
fully describe such long-range degree correlations (LRDCs)
between degrees $k$ and $k'$ of two nodes separated by $l$ in
the sense of the shortest path length, we introduced the joint
probability $P(k,k',l)$ and four conditional probabilities
$P(l|k,k')$, $P(k'|k,l)$, $P(k,k'|l)$, and $P(k',l|k)$. These
probabilities are not independent, and several relations
between them have been presented with the aid of the Bayes'
theorem. It has also been shown that the above probability
functions include the probabilities $P_{\text{nn}}(k,k')$ and
$P_{\text{nn}}(k'|k)$ describing nearest neighbor degree
correlations as a special case. Furthermore, we have
analytically calculated these five probabilities for a network
without any degree correlations at an arbitrary distance under
the local tree assumption and the mean-field approximation. The
results for Erd\H{o}s-R\'{e}nyi random graphs and scale-free
random networks agree well with numerical ones. The
probabilities for long-range uncorrelated networks enable us to
judge the existence of LRDCs in a given network and capture the
feature of correlations. Finally, we analyzed LRDCs in
real-world networks within the present framework and found that
the coauthorship network possesses LRDCs in which small degree
nodes strongly repel each other.

Although we have just prepared tools for analyzing LRDCs,
it is quite interesting to study relations between LRDCs
and many network properties such as the robustness of a
network, fractality, synchronization, to name a few. Our joint and
conditional probabilities are three-variable functions and
are not easy to handle. Thus, it is also important to develop
intuitive indices characterizing LRDCs, like a measure of
the strength of the repulsive correlation between similar
degree nodes, on the basis of these probabilities.

\begin{acknowledgements}
The authors thank S.~Mizutaka and T.~Hasegawa for fruitful
discussions. This work was supported by a Grant-in-Aid for
Scientific Research (No.~16K05466) from the Japan Society
for the Promotion of Science. T.T. acknowledges
the financial support through JST ERATO Grant Number JPMJER1201, Japan.
\end{acknowledgements}


\begin{thebibliography}{99}
\bibitem{comment2} While Ref.~\cite{Melnik16} calculates the distance distribution for a network consisting of a connected component, our treatment covers networks composed of multiple disconnected components.
\end{thebibliography}
\end{document}